\documentclass[twocolumn,times]{aastex62}
\usepackage{gensymb}
\usepackage{amsmath}
\usepackage{amssymb}
\usepackage{textcomp}

\pdfoutput=1

\received{September 13, 2018}
\revised{February 01, 2019}
\accepted{February 02, 2019}
\submitjournal{ApJ}

\shorttitle{NGFS:~V.~Discovery of an isolated dwarf-dwarf galaxy pair at $z=0.3$ }
\shortauthors{Evelyn J. Johnston et al.}

\begin{document}

\title{The Next Generation Fornax Survey (NGFS):~V.~Discovery of a dwarf-dwarf galaxy pair at $z=0.30$ and its characterization using deep VLT/MUSE observations}

\correspondingauthor{Evelyn J. Johnston}
\email{ejohnston@astro.puc.cl}

\author[0000-0002-2368-6469]{Evelyn J. Johnston}
\altaffiliation{FONDECYT Postdoctoral Fellow}
\affiliation{Institute of Astrophysics, Pontificia Universidad Cat\'olica de Chile, Av.~Vicu\~na Mackenna 4860, 7820436 Macul, Santiago, Chile}

\author[0000-0001-8654-0101]{Paul~Eigenthaler}
\altaffiliation{CASSACA Postdoctoral Fellow}
\affiliation{Institute of Astrophysics, Pontificia Universidad Cat\'olica de Chile, Av.~Vicu\~na Mackenna 4860, 7820436 Macul, Santiago, Chile}

\author[0000-0003-0350-7061]{Thomas H.~Puzia}
\affiliation{Institute of Astrophysics, Pontificia Universidad Cat\'olica de Chile, Av.~Vicu\~na Mackenna 4860, 7820436 Macul, Santiago, Chile}

\author[0000-0001-7966-7606]{Yasna Ordenes-Brice\~no}
\altaffiliation{PUC-HD Graduate Student Exchange Fellow}
\affiliation{Institute of Astrophysics, Pontificia Universidad Cat\'olica de Chile, Av.~Vicu\~na Mackenna 4860, 7820436 Macul, Santiago, Chile}
\affiliation{Astronomisches Rechen-Institut, Zentrum f\"ur Astronomie der Universit\"at Heidelberg, M\"onchhofstra{\ss}e 12-14, D-69120 Heidelberg, Germany}

\author[0000-0003-3009-4928]{Matthew A.~Taylor}
\altaffiliation{Gemini Science Fellow}
\affiliation{Gemini Observatory, Northern Operations Center, 670 North A'ohoku Place, Hilo, HI 96720, USA}

\author[0000-0002-5897-7813]{Karla~Alamo-Mart\'inez}
\altaffiliation{FONDECYT Postdoctoral Fellow}
\affiliation{Departamento de Astronomia, Instituto de F\'isica, Universidade Federal do Rio Grande do Sul, 91501-970 Porto Alegre, R.S, Brazil}

\author[0000-0003-1184-8114]{Patrick C{\^o}t{\'e}}
\affiliation{NRC Herzberg Astronomy and Astrophysics, 5071 West Saanich Road, Victoria, BC V9E 2E7, Canada}

\author[0000-0002-8835-0739]{Gaspar Galaz}
\affiliation{Institute of Astrophysics, Pontificia Universidad Cat\'olica de Chile, Av.~Vicu\~na Mackenna 4860, 7820436 Macul, Santiago, Chile}

\author[0000-0002-1891-3794]{Eva K.\ Grebel}
\affiliation{Astronomisches Rechen-Institut, Zentrum f\"ur Astronomie der Universit\"at Heidelberg, M\"onchhofstra{\ss}e 12-14, D-69120 Heidelberg, Germany}

\author[0000-0002-2363-5522]{Michael Hilker}
\affiliation{European Southern Observatory, Karl-Schwarzchild-Str. 2, D-85748 Garching, Germany}

\author[0000-0002-7214-8296]{Ariane~Lan\c{c}on} 
\affiliation{Observatoire astronomique de Strasbourg, Universit\'e de Strasbourg, CNRS, UMR 7550, 11 rue de l'Universite, F-67000 Strasbourg, France}

\author[0000-0003-4197-4621]{Steffen Mieske}
\affiliation{European Southern Observatory, 3107 Alonso de C\'ordova, Vitacura, Santiago}

\author[0000-0003-4945-0056]{Ruben S\'anchez-Janssen}
\affiliation{STFC UK Astronomy Technology Centre, Royal Observatory, Blackford Hill, Edinburgh, EH9 3HJ, UK}

\author[0000-0002-2204-6558]{Yu Rong}
\altaffiliation{CASSACA Postdoctoral Fellow}
\affiliation{Institute of Astrophysics, Pontificia Universidad Cat\'olica de Chile, Av.~Vicu\~na Mackenna 4860, 7820436 Macul, Santiago, Chile}

\begin{abstract}
We report the detection of a pair of dwarf galaxies at $z\!=\!0.30$ which may be in the early stages of an interaction. Both galaxies have stellar masses of $<10^{9}\,M_\odot$, and display a projected separation of $\sim\!29$\,kpc and a physical separation of $\sim\!240$\,kpc. Evidence of ongoing star formation has been found in both galaxies, with neither one showing an enhanced star-formation rate that would be expected if they were already interacting. One galaxy displays a disturbed morphology but shows ordered gas rotation, which may reflect a previous minor merger event in the recent history of that system. The nearest spectroscopically confirmed neighbour lies at a distance of 38~Mpc. These results indicate that these dwarf galaxies have no neighbouring massive galaxies, however with the data available we have been unable to determine whether these galaxies are isolated in the field or belong to a group of low-mass galaxies.~As a detection of a rare dwarf-dwarf pair beyond the Local Universe, this system provides an uncommon opportunity to explore the properties of galaxy groups in the low-galaxy mass regime as a function of redshift. 
\end{abstract}

\keywords{galaxies: dwarf  --- galaxies: star formation --- galaxies: evolution}

\section{Introduction}\label{sec:intro}
According to the $\Lambda$CDM model, galaxies have been built up through hierarchical formation, forming through mergers and the accretion of smaller galaxies \citep[e.g.][]{Guo_2008}.~This theory extends to galaxies of all masses, and suggests that dwarf galaxies, with stellar masses ${\cal M}_*<\!10^9M_\sun$, dominate the universe in terms of numbers  \citep{Marzke_1997,fer16, bah17} and are involved in the majority of merger events throughout the cosmic evolution \citep[e.g.][]{del06, mis16}.~While simulations have shown that a substantial fraction of stellar mass within galaxies has formed in situ \citep[e.g.][]{Pillepich_2015}, the majority of the stellar content in the brightest galaxies today formed at high redshifts ($z\gtrsim5$) in low-mass systems, but was assembled at lower redshifts \citep[$z\lesssim1$; e.g.][]{del07}.~Because of this, understanding the dynamics and stellar population properties of galaxy-galaxy mergers in the dwarf mass regime as a function of redshift is a crucial, but an inherently difficult task.

Studies of nearby gas-rich dwarf galaxies have revealed that their star-formation rates are consistent with the scenario of continuous or extended star formation with amplitude variations \citep{Tosi_1991, Greggio_1993, Grebel_1997, Grebel_1999, Weisz_2011}. However, the  triggers of these short periods of enhanced star formation are still unknown, with explanations including secular processes \citep{Brosch_2004}, gas inflow via filaments \citep{Sanchez_2015} and mergers, which would induce star-formation throughout the whole galaxy, leading to the formation of Blue Compact Dwarf galaxies \citep[BCDs,][]{Pustilnik_2001, Bekki_2008, Lelli_2012a, Lelli_2012b, Koleva_2014} and isolated dwarf galaxies with ongoing star formation \citep{Telles_1995,Noeske_2011}.

Further evidence of mergers and accretion includes disturbed kinematics, such as those observed in the halo of NGC~4449 \citep{Hunter_1998} and in the dwarf spheroidal galaxy Andromeda {\sc ii} \citep{Amorisco_2014}, and faint structures within the galaxy, such as the shells detected by \citet{Paudel_2017} in three non-star-forming dwarf galaxies in the Virgo Cluster, which are suggestive of a dry-merger scenario. In the case of  the Sextans dSph, \citet{Cicuendez_2018} found evidence of both kinematic anomalies and different distributions for the metal-rich and metal-poor stars, and DDO~68 has been found to contain faint stellar streams resulting from the accretion of two dwarf companions \citep{Annibali_2016} and a faint companion detected in HI \citep{Cannon_2014}. A faint companion has also been detected for NGC~4449  \citep[NGC~4449B;][]{Martinez_2012, Rich_2012},  and \citet{Koch_2015} have detected two objects  at $z=0.12$ with different colors that are possibly linked by a tidal tail, which is thought to be evidence of an ongoing dwarf-dwarf merger. 

Applying such studies beyond the Local Group is difficult due to the faintness of the dwarf galaxies.~However, various studies have characterized the properties of SDSS-selected dwarf-dwarf pairs in the nearby universe out to $\sim\!40$\,Mpc \citep[$z\!\simeq\!0.0091$;][]{Stierwalt_2015, Pearson_2016, Privon_2017}.~In an effort to measure the properties of dwarf galaxies over a range of redshifts in a statistically meaningful way, several deep, multi-waveband surveys are now being carried out to search for these faint galaxies.~Examples include the Solitary Local dwarfs survey \citep[SoLo, $<$~3Mpc;][]{Higgs_2016}, the Next Generation Virgo Cluster Survey \citep[NGVS, $z\approx0.004$;][]{Ferrarese_2012}, the Next Generation Fornax Survey \citep[NGFS, $z\approx0.005$;][]{Munoz_2015, Ordenes_2018a}, the dwarf galaxies in the Coma Cluster survey \citep[$z\approx0.02$;][]{Kourkchi_2012} and Abell~85 \citep[$z\approx0.55$;][]{Agulli_2014}.
However, the majority of these surveys  search for dwarf galaxies in dense environments, leading to an environmental bias in studies of dwarf galaxies beyond the Local Group. Consequently, the detection of isolated dwarf galaxies beyond the Local Group is important to help us build up a clearer picture of dwarf galaxy evolution and the triggers for their enhanced star formation in all environments.

In this paper, we describe the discovery and characterization of a pair of  dwarf galaxies in the field at $z\simeq0.30$ behind the Fornax Cluster.~The galaxies  show no evidence of direct interaction, and it is likely that they are still on their first or second approach.~The detection of interacting dwarf galaxies in the field is a rare occurrence, making the characterization of this system particularly significant in terms of our understanding of dwarf galaxy evolution as it shows that even in isolated environments, the role of galaxy mergers and interactions cannot be ruled out completely \citep{Stierwalt_2017}.

This paper is laid out as follows.~Section~\ref{sec:DR} describes the data sets used for the analysis and their reduction, while Section~\ref{sec:analysis} outlines the various steps in our analysis, and Section~\ref{sec:conclusions} discusses the results and summarises our conclusions. 
Throughout this paper we assume a Hubble constant of H$_\text{o}=70\pm1.3$~km~s$^{-1}$~Mpc$^{-1}$ \citep{Carroll_1992}.

\begin{figure*}[!t]
\begin{center}
\includegraphics[scale=0.4,angle=0,width=\linewidth]{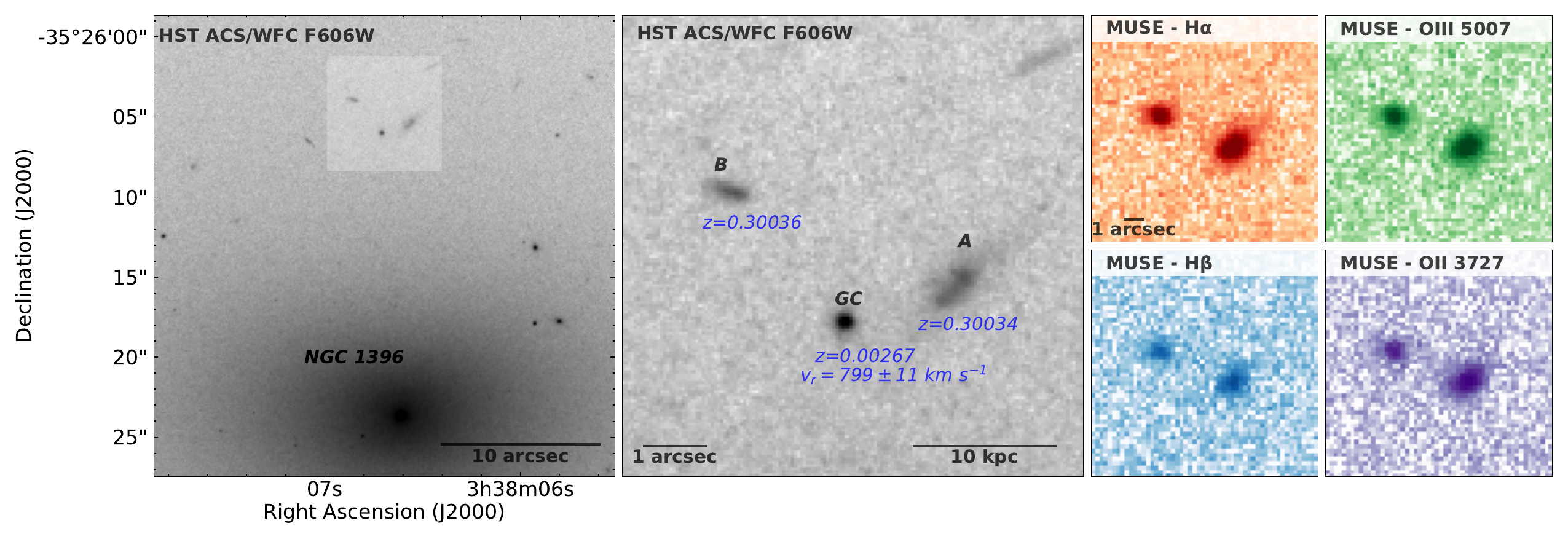}
\caption{({\it Left}): Hubble Space Telescope ACS/WFC F606W-filter image of NGC\,1396 with the pair of background galaxies highlighted. ({\it Center}): A zoom-in on the highlighted area, showing the pair of background galaxies (${\cal A}$ and ${\cal B}$) and a Fornax galaxy globular cluster (GC). The redshifts of the three objects are marked on the image, where the redshifts of the galaxies have been measured from the gas emission lines while that of the globular cluster was measured from the stellar absorption lines.~({\it Right}): The same zoom-in area from the middle panel in the MUSE datacube, showing, clockwise from top-left, the continuum-subtracted H$\alpha$, [O{\sc iii}]$_{5007}$, [O{\sc ii}]$_{3727}$ and H$\beta$ emission maps.
 \label{fig:images}}
\end{center}
\end{figure*}

\section{Observations}\label{sec:DR}
An F606W-band image from the Hubble Space Telescope ACS/WFC is presented in Figure~\ref{fig:images}, showing the location of the dwarf galaxy pair relative to NGC\,1396, a dwarf elliptical galaxy in the Fornax Cluster. This image was observed as part of proposal ID 10129 (PI: Puzia), with a total exposure time of 2108s. A zoom-in on the system  demonstrates the irregular morphology of Galaxy~${\cal A}$, which is the brighter of the two galaxies, with bright knots that may be due to short-lived periods of enhanced star formation, and a faint tidal tail extending North-West from the galaxy. This disturbed morphology is reminiscent of NGC7796-DW1, the early type dwarf companion to the isolated elliptical NGC~7796 that is thought to have accreted cold, metal-rich gas from NGC~7796 \citep{Richtler_2019}. No obvious bridge can be seen connecting Galaxies~${\cal A}$ and ${\cal B}$, and no other companions lie within the field-of-view.~Due to seeing effects on the ground based data, these features are not evident in the data used throughout this study, although the light profiles of both galaxies are visibly different from that of the nearby globular cluster also marked in the zoom-in image.

\subsection{NGFS data}\label{sec:DR_NGFS}
The imaging data in this paper is from the NGFS, which is an ongoing, multi-waveband survey of the central 30~deg$^2$ of the Fornax Cluster using Blanco/DECam for near-ultraviolet/optical and VISTA/VIRCAM for near-infrared photometry.~The current NGFS field is built up by nine DECam tiles centered on NGC\,1399, and the target dwarf galaxies of this study lie within the central tile. Below we provide relevant details of the NGFS data, while a more detailed overview of the NGFS observations and data reduction can be found in \citet{Eigenthaler_2018} and \cite{Ordenes_2018a}.

The NGFS data used in this paper consisted of the DECam $u^\prime$, $g^\prime$, $i^\prime$ and $K_s^\prime$ photometry of the central tile.~The exposure times were determined to obtain a signal-to-noise ratio (S/N) of 5 at 26.5, 26.1, 25.3 AB and 23.3 mag in the  $u^\prime$-, $g^\prime$-, $i^\prime$-, and $K_s^\prime$-band, respectively. The data reduction was carried out in two steps: The DECam Community Pipeline \citep[CP,][]{Valdez_2014} first applies the bias correction, flat-fielding and cross-talk correction, and further processing was carried out using the \textsc{Astromatic} software, which calculated the astrometric calibration, stacked the individual images, and  applied the flux calibration using \textsc{scamp}, \textsc{swarp} and \textsc{source extractor} \citep{Bertin_1996, Bertin_2002, Bertin_2006}.

\subsection{MUSE data}\label{sec:DR_MUSE}
The spectroscopic data used in this paper was observed with the Multi-Unit Spectroscopic Explorer \citep[MUSE;][]{Bacon_2010} at the Very Large Telescope (VLT). MUSE  is an optical wide-field integral-field spectrograph with a field of view of $1\arcmin\times1\arcmin$, a spatial resolution of 0.2\arcsec/pixel, and a spectral resolving power ranging from \textit{R}$\simeq$1770 at 4800\,\AA\ to \textit{R}$\simeq$3590 at 9300\,\AA.

MUSE observations of the dwarf system are publicly available on the ESO Science Archive Facility as part of program 094.B-0895, for which the target was the nearby dwarf galaxy NGC\,1396 \citep{men16}.~The observations were carried out between December 16, 2014 and January 21, 2015, and consist of 16$\times$1300\,second exposures covering a field of view of $2\arcmin\times 1\arcmin$, with a small offset and $90\degree$ rotation between each pair of exposures.~Separate sky exposures were taken for each pair of science exposures, and standard stars were observed for each night of observations. Additional sky flats were obtained within 7 days, and internal lamp flats were taken with the calibration lamps at the start of each set of observations to account for the time-dependent, temperature-related variations in the flux levels between each IFU. 

The data were reduced using the ESO MUSE pipeline \citep[v3.12.3,][]{Weilbacher_2012} in the ESO Recipe Execution Tool (EsoRex) environment \citep{ESOREX}.~The associated raw calibrations were used to create master bias and flat field frames for each night of observations, which were applied to the raw science and standard-star observations as part of the pre-processing steps. A flux calibration solution was obtained for each night using the standard-star exposures, and the sky continuum was measured for each pair of science exposures. The post-processing step of the data reduction applied the flux calibration and sky subtraction to the science exposures, and the data were  combined to produce the final datacube. As a final step, the residual sky contamination was cleaned using the Zurich Atmosphere Purge code \citep[ZAP, ][]{Soto_2016}, in the ESO Recipe Flexible Execution Workbench environment \citep[EsoReflex,][]{Freudling_2013}. The continuum-subtracted emission line maps for H$\alpha$, H$\beta$, [O{\sc ii}]$_{3727}$ and [O{\sc iii}]$_{5007}$ are presented in Figure~\ref{fig:images} to give a comparison between the HST data and the ground based data.


\begin{figure*}
\begin{center}
\includegraphics[angle=0,width=0.99\linewidth]{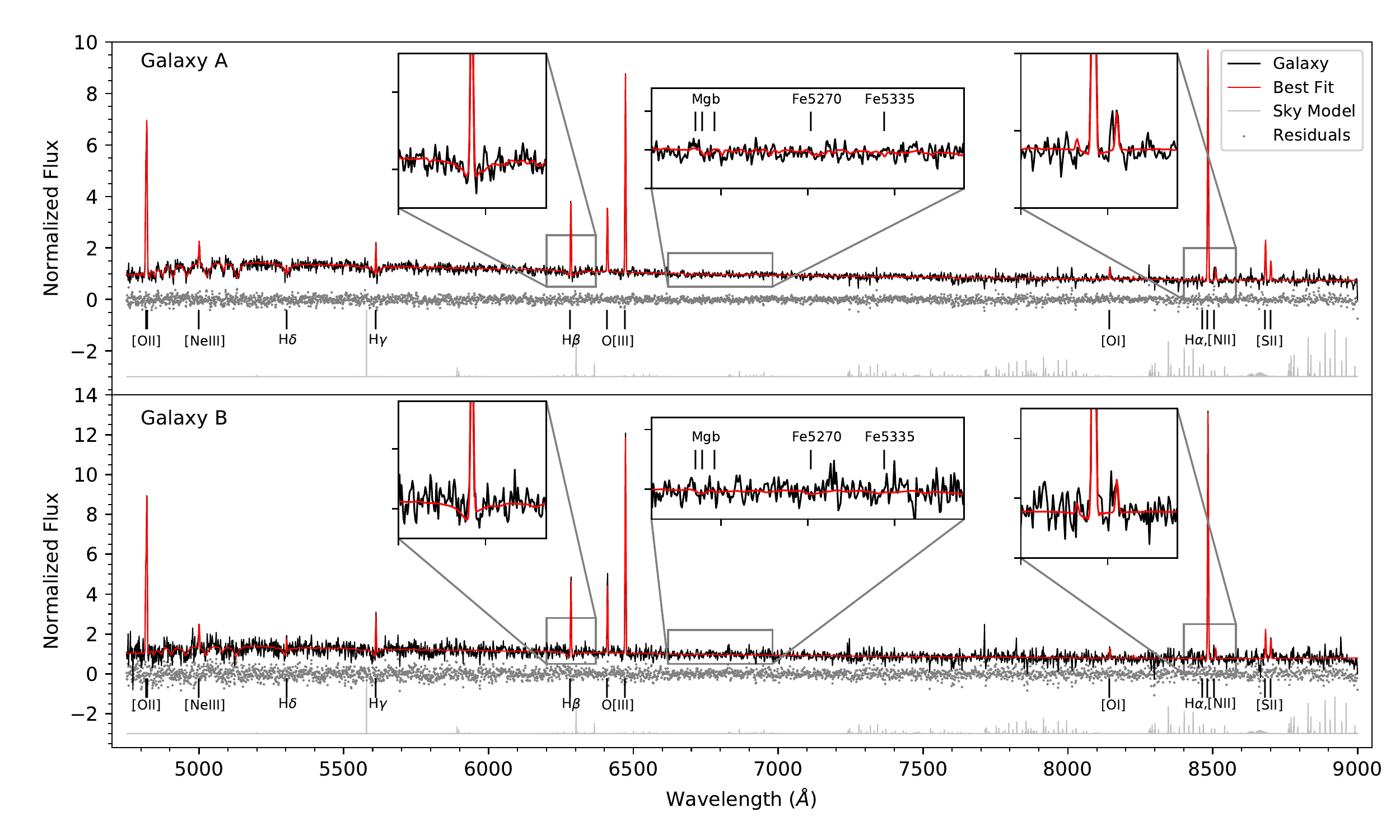}
\caption{The pPXF fits to the spectra for Galaxy~${\cal A}$ (\textit{top}) and ${\cal B}$ (\textit{bottom}), showing the input spectrum in black, the best-fitting model spectrum from the combined fits to the stars and gas in red, and the residuals as grey points. The grey line shows an example sky spectrum for reference, and the insets show a zoom-in on certain emission and absorption features.
}
\label{fig:fits}
\end{center}
\end{figure*}

\section{Analysis}\label{sec:analysis}
\subsection{Determination of the redshift}\label{sec:redshift}
An integrated spectrum was extracted from the MUSE datacube for each galaxy using a circular aperture of radius 0.8\arcsec\ and 0.6\arcsec\ centered on galaxies ${\cal A}$ and ${\cal B}$, respectively, where these apertures were selected to optimise the S/N. Due to the faintness of these galaxies, the extracted spectra have S/N of $\sim\!11$ and $\sim\!6$ per~\AA\ for galaxies~${\cal A}$ and ${\cal B}$, respectively, as measured in the continuum at $\sim6000$\AA.

The gas and stellar kinematics of each galaxy were measured using the penalized Pixel Fitting (pPXF) code of \citet{Cappellari_2004}, which combines stellar templates of defined ages and metallicities with line-of-sight velocity distributions and velocity dispersions to obtain a model spectrum that best fits the galaxy spectrum.~Both galaxies were fitted using a two-component fit, where the primary component modeled the continuum and absorption features to extract the stellar kinematics while the secondary component modeled the emission lines to measure the gas kinematics.~The MILES stellar library \citep{Sanchez_2006} was used for the primary component, using 300 template spectra with an age range of  0.06 to 18\,Gyrs and a range of metallicity between [Fe/H]~$\!\approx\!-1.71$ and $+0.22$\,dex.~For the secondary component a series of 14 emission-line templates was created, where each template represented one of the prominent emission features between [O{\sc ii}]$_{3726\text{\AA}}$ and [S{\sc ii}]$_{6731\text{\AA}}$ and used the same wavelength range and resolution as the stellar template spectra. By modeling the emission lines in this way, the strength of each emission line was allowed to vary to best fit the galaxy spectrum, while the radial velocity and velocity dispersion were held constant across all the templates.~All the template spectra were  convolved with low-order additive polynomials to model the flux calibration of the continuum and reduce mismatched templates selected.~An illustration of the best fit to the data can be seen in Figure~\ref{fig:fits} for both dwarf galaxies.

\begin{figure*}
\begin{center}
\includegraphics[angle=0,width=0.9\linewidth]{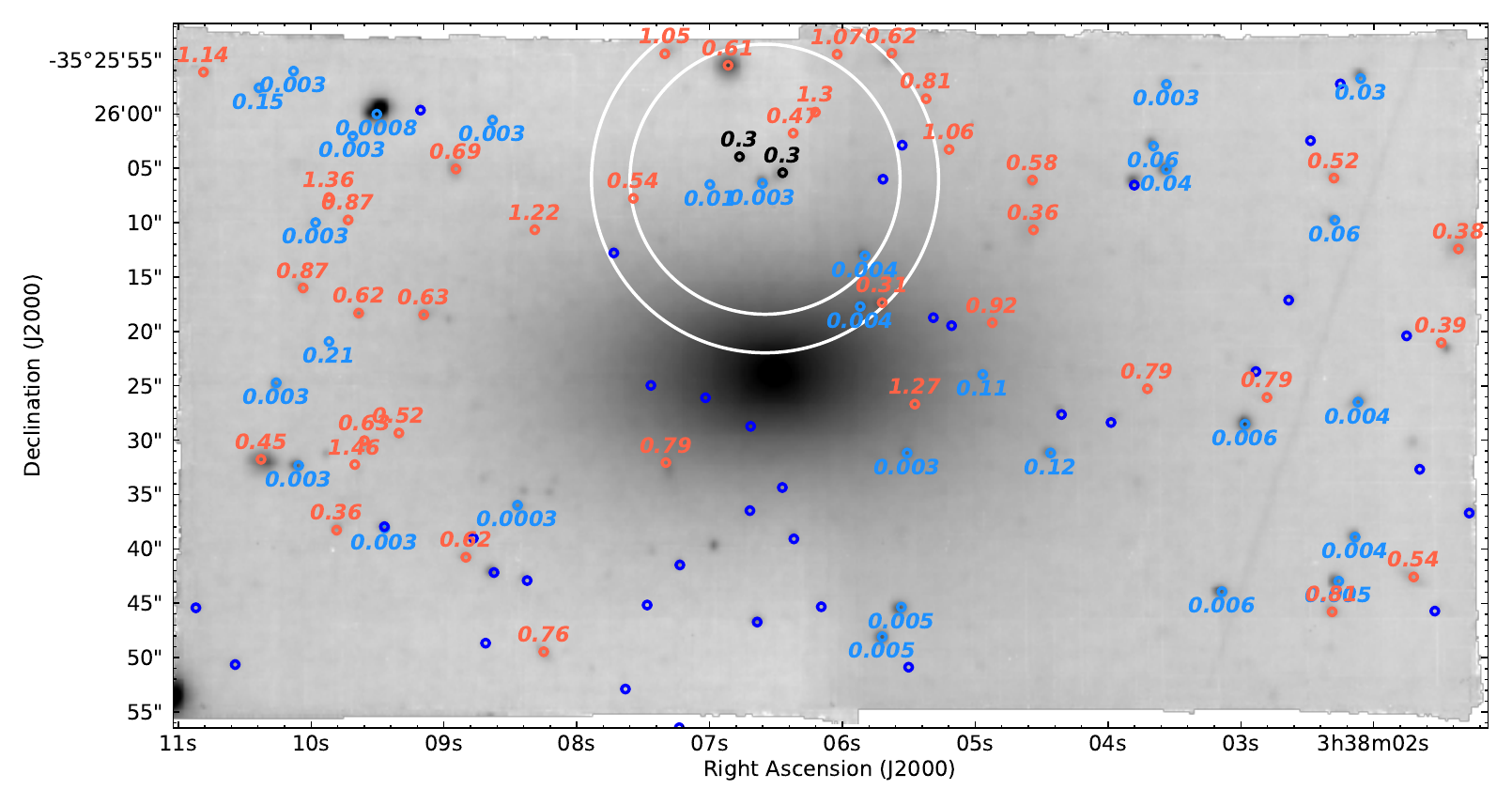}
\caption{The $i^{\prime}$-band image of NGC~1396 created from the MUSE datacube, with the redshifts overplotted for all objects for which a redshift could be measured. Red and light blue circles and labels represent galaxies at $z>0.30$ and $z<0.30$ respectively, while darker blue circles reflect globular clusters at the same redshift as NGC~1396 from \citet{men16} for which the S/N was too low to obtain a reliable redshift with the MUSE data. The concentric white circles are centered on the dwarf-dwarf pair at $z=0.30$, and display the circular field-of-view within which we can fully search for neighbours (inner circle) and the circular distance to the nearest neighbour (outer circle). The image is orientated with North to the top and East to the left.}
\label{fig:redshifts}
\end{center}
\end{figure*}

The results for the gas kinematics analysis for both dwarfs are given in Table~\ref{tab:kinematics}, where the uncertainties have been estimated through a series of Monte Carlo simulations on the model spectrum, which was obtained as the best fitting combination of templates to the galaxy spectra, with simulated levels of noise to match the S/N of the original spectrum. For completeness, the spectrum of the nearby candidate globular cluster was extracted and the stellar kinematics measured in the same way, which confirmed that it lies at the same redshift as NGC~1396 \citep{Drinkwater_2001}. It should be noted at this point that the velocity dispersion of both galaxies and the globular cluster are smaller than the instrumental resolution. While the recent modifications to pPXF, as outlined in \citet{Cappellari_2017}, should give reliable measurements for such small velocity dispersions, care should be taken when using these values.

It can be seen that the gas velocities of the two galaxies are in almost-perfect agreement with one another, differing by  less than 10~km/s, considering the uncertainties, at a redshift of $z\!\simeq\!0.30$, indicating that their proximity on sky is not simply a projection effect. The galaxies have a projected separation on sky of $3.82\pm0.06\arcsec$, which corresponds to a separation of $28.8\pm0.5$~kpc at a redshift of $z=0.30034$. Taking into account the differences in the redshifts and the uncertainties on these measurements, the physical separation of the two galaxies becomes $242\pm339$~kpc. This separation is consistent with 0, and even  the maximum separation within the uncertainty is still less that the separation criteria used by \citet{Stierwalt_2015} to identify dwarf galaxy companions in the \textsc{TiNy Titans} survey.

The stellar kinematics of the two galaxies are not included in Table~\ref{tab:kinematics} since the low S/N of the spectra led to artificial broadening of the stellar velocity dispersion and unreliable measurements for the stellar kinematics.
The effect of this low S/N on the fits can be seen in the insets in Figure~\ref{fig:fits}, especially in the case of Galaxy~${\cal B}$.~A model sky spectrum is shown for reference in the plot, which was created using the Cerro Paranal Sky Model \citep{Noll_2012, Jones_2013}.~It can be seen that the majority of the noise in the spectrum at redder wavelengths corresponds to the presence of sky line residuals that were not removed completely during the sky subtraction steps.

\subsection{Exploring the environment}\label{sec:environment}
\citet{Besla_2018} found that the majority of systems containing multiple dwarf galaxies within the redshift range of \mbox{$0.013<z<0.0252$} are pairs, with triples and higher order multiples being rare. However, such groups containing up to 5 dwarf galaxies have been found to exist out to $z=0.04$ by \citet{Stierwalt_2017}. Therefore, it is important to determine whether these galaxies are an isolated pair or members of a group.

\begin{figure}
\begin{center}
\includegraphics[angle=0,width=0.99\linewidth]{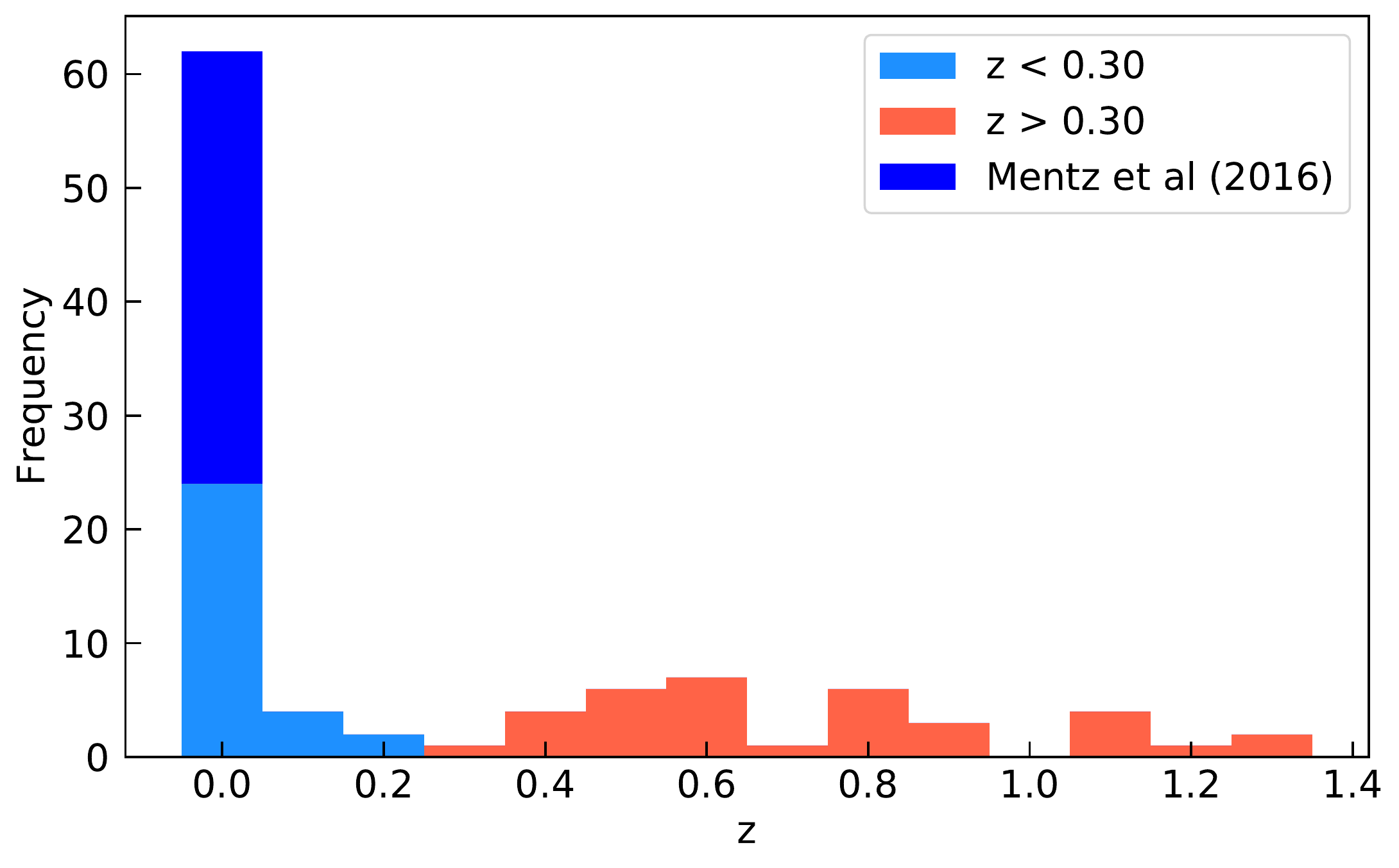}
\caption{A histogram showing the distribution of spectroscopic redshifts for all targets shown in Fig.~\ref{fig:redshifts}. The globular clusters from \citet{men16}, for which redshifts could not be obtained with the MUSE data, were assigned a redshift of $z=0.003$ to match NGC~1396.}
\label{fig:histogram}
\end{center}
\end{figure}

As a first approach, the analysis described in Section~\ref{sec:redshift} was applied to all other faint sources detected within the datacube to look for neighbouring galaxies at the same redshift. These sources and their redshifts are shown in Fig.~\ref{fig:redshifts}, and the redshift distribution is given in Fig.~\ref{fig:histogram}. Evidence of large-scale structure behind NGC~1396 can be seen in these plots, with evidence of objects at redshifts out to 1.3. This finding agrees with the work of \citet{Hilker_1999} and \citet{Drinkwater_2000}, who found evidence of a large-scale structure filament of galaxies extending out to $z=0.3$ behind the Fornax Cluster. Since the dwarf-dwarf pair lies offset from the centre of the field, it is only possible to carry out a complete search for neighbours within a circular volume of radius of $\sim12$\arcsec\ on sky, as marked by the inner white circle in Fig.~\ref{fig:redshifts} and which corresponds to $\sim90$kpc at $z=0.3$. However, within the full field-of-view available with the MUSE data, the nearest neighbour in terms of redshift lies $\sim16$\arcsec\ to the south-west (distance marked by the outer white circle) at a redshift of $z=0.308$. This distance and angle correspond to a physical (proper) distance of 49.3~Mpc from the dwarf-dwarf pair, indicating that this nearest neighbour is not part of the same group if we use a maximum projected separation of 1.5~Mpc for the nearby dwarf galaxy groups in \citet{Stierwalt_2017} as a guide.

\begin{figure}
\begin{center}
\includegraphics[angle=0,width=0.99\linewidth]{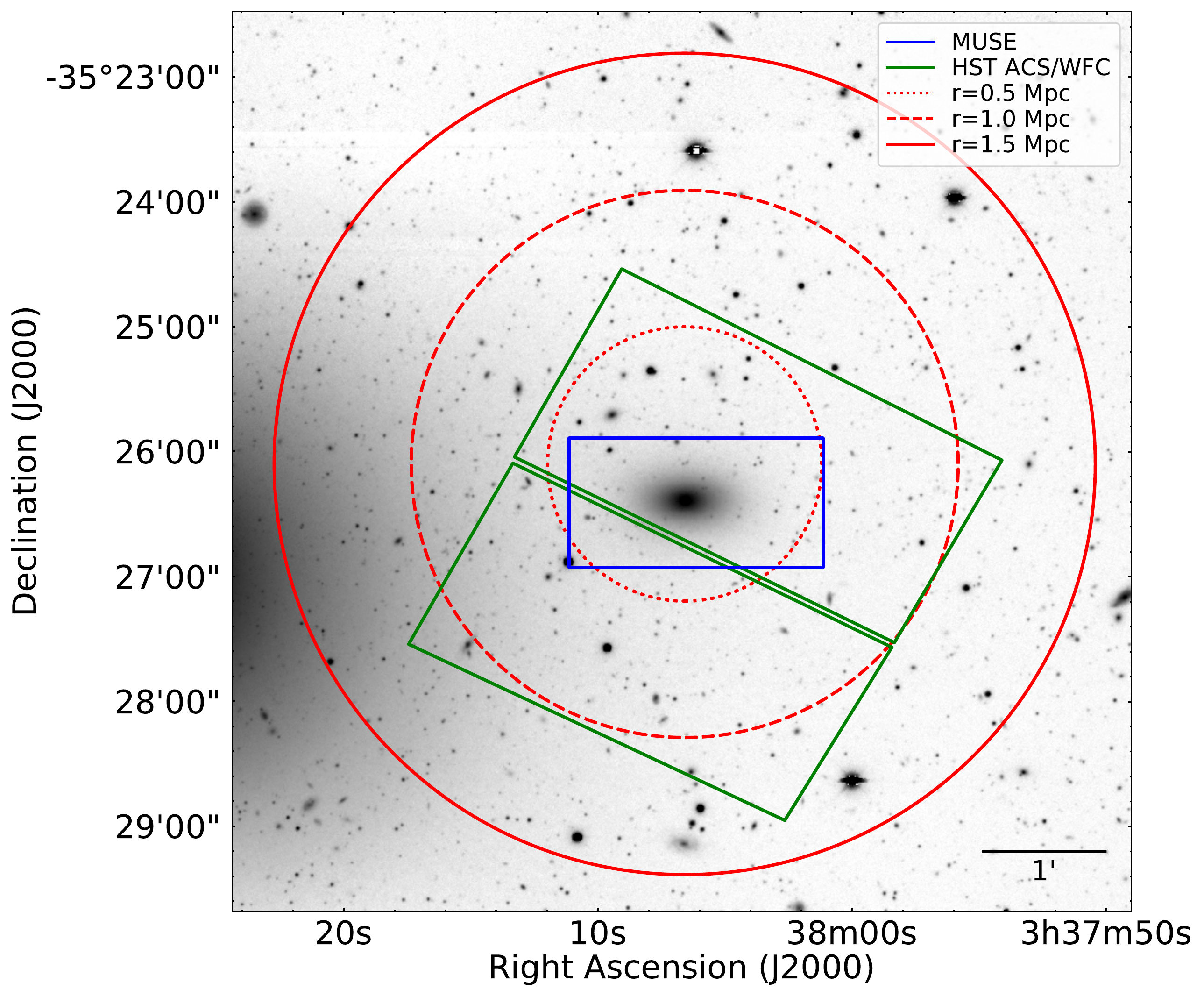}
\caption{The NGFS $i^{\prime}$-band image centered on the dwarf-dwarf galaxy pair. The blue and green rectangles represent the fields-of-view of the MUSE and HST ACS/WFC data used in this study respectively, while the dotted, dashed and solid red lines reflect distances of 0.5, 1.0 and 1.5~Mpc on sky at $z=0.3$. }
\label{fig:fields}
\end{center}
\end{figure}

In order to better determine the environment, we turn to the literature to obtain spectroscopic redshifts for objects within a volume of radius 1.5~Mpc. Figure~\ref{fig:fields} presents this search area centered on the dwarf-dwarf pair from the NGFS data as concentric circles marking radii of 0.5, 1.0 and 1.5~Mpc at a redshift of $z=0.3$. The fields of view of the MUSE and HST ACS/WFC data are also superimposed for reference. Spectroscopic redshifts for objects in the direction of Fornax were taken from the Fornax Cluster Spectroscopic Survey \citep[FCSS,][]{Drinkwater_2000, Morris_2007} and the 2dF Galaxy Redshift Survey \citep{Colless_2003}, and the physical separations were measured relative to the dwarf-dwarf pair. Since both these surveys focus on targets brighter than 10 (FCSS) and 13~mag (2DF) in the \textit{b$_j$} band, and do not include the dwarf-dwarf pair, these catalogs have been used to identify whether the dwarf-dwarf pair has a more massive companion. From these data sets, the nearest neighbour was found in the FCSS point source catalog at a redshift of $\sim0.294$ and a physical distance of $\sim38$~Mpc. Therefore, we have found no massive neighbours within 1.5~Mpc of the dwarf-dwarf pair. 

In order to probe the environment to fainter magnitudes, one can use the fluxes from imaging data to determine the photometric redshifts of targets within the field of view. The NGFS data of this field currently only contains the $u^\prime$-, $g^\prime$-, $i^\prime$- and $K_s$-bands, which are not ideal for such studies since the low number of filters and the large gap in wavelength between the $i^\prime$- and $K_s$-bands lead to poor SED fits with the templates. Before applying this analysis to the entire 1.5~Mpc volume, an initial test was carried out within the field-of-view of the MUSE data to allow a direct comparison with the spectroscopic redshifts shown in Fig.~\ref{fig:redshifts}. The flux of each object was measured in each of the NGFS images using aperture photometry with apertures of radii 6, 8, 10 and 12 pixels (1.7, 2.3, 2.9 and 3.5~arcseconds), and their photometric redshifts measured using the EAZY photometric redshift code \citep{Brammer_2008}. However, when the photometric redshifts were compared to the spectroscopic redshifts, the results were found to be inconsistent between the two measurements for all apertures, and the uncertainties were very large, with mean and median uncertainties in photometric redshift of the order 0.12 and 0.23. As a consequence, this analysis with the NGFS images was considered too unreliable to apply to the rest of the field.

As a result of the analysis presented in this section, we have been able to rule out the possibility of a more massive companion to the dwarf-dwarf pair. However, with the data available, we have been unable to determine whether these galaxies lie isolated in the field or are part of a group of low-mass galaxies, such as those detected in the Local Universe by \citet{Tully_2006} and \citet{Stierwalt_2017}.

\begin{deluxetable*}{lhrrr}
\tablenum{1}
\tablecaption{Kinematics measurements and physical parameters for the interacting dwarf galaxies and the foreground globular cluster (see center panel of Fig.~\ref{fig:images}). \label{tab:kinematics}}
\tablewidth{0pt}
\tablehead{
\colhead{ }  & \nocolhead{ } & \colhead{Galaxy ${\cal A}$} & \colhead{Galaxy ${\cal B}$} & \colhead{GC} 
}
\startdata
Ra & & 03:38:06.47 & 03:38:06.75 & 03:38:06.60  \\
Dec & & $-$35:26:05.78 & $-$35:26:04.38 & $-$35:26:06.38  \\
$v$ (km/s) & & $76967\pm5$ & $76975\pm6$ & $799\pm11$ \\
$\sigma$ (km/s) & & $19\pm6$ & $11\pm9$ & $29\pm20$  \\
$z$  & & $0.30034\pm0.00002$ & $0.30036\pm0.00003$ & $0.00267\pm0.00004$  \\
Distance (Mpc)  & & $1554.76\pm0.14$ & $1555.0\pm0.2$ & $11.6\pm0.2$  \\
\\
$m_{g^\prime}$ (mag) & & $23.04\pm0.12$ & $24.44\pm0.05$ & $23.38\pm0.02$  \\
$m_{i^\prime}$ (mag) & & $22.2\pm0.3$ & $23.60\pm0.16$ & $22.56\pm0.02$  \\
$M_{g^\prime}$ (mag) & & $-17.94\pm0.12$ & $-16.55\pm0.05$ & $-7.71\pm0.02$  \\
$M_{i^\prime}$ (mag) & & $-18.5\pm0.3$ & $-17.14\pm0.16$ & $-8.53\pm0.02$  \\
$\langle\mu_{g^\prime}\rangle_e$ (mag/arcsec$^2$) & & $21.0\pm0.5$ & $21.0\pm0.4$ & $-$  \\
$\langle\mu_{i^\prime}\rangle_e$ (mag/arcsec$^2$) & & $20.1\pm0.7$ & $21.1\pm0.8$ & $-$  \\
$R_{e,g^\prime}$ ($\arcsec$) & & $0.37\pm0.08$ & $0.24\pm0.04$ & $-$  \\
$R_{e,i^\prime}$ ($\arcsec$) & & $0.39\pm0.08$ & $0.19\pm0.06$ & $-$  \\
$R_{e,g^\prime}$ (kpc) & & $2.0\pm0.2$ & $1.3\pm0.2$ & $-$  \\
$R_{e,i^\prime}$ (kpc) & & $2.1\pm0.4$ & $1.0\pm0.3$ & $-$  \vspace{1mm} \\
\\
$\log_{10}({\cal M}_*/M_{\sun}$) & & $8.8\pm0.3$ & $8.2\pm0.2$ & $-$  \\
E(B-V) (mag) & & $0.06\pm0.02$ & $0.10\pm0.03$& $-$ \\
SFR$_{\text{H}\alpha}$ (M$_{\sun}$ yr$^{-1}$) & & $0.083\pm0.009$ & $0.059\pm0.010$ & $-$ \\
$\log_{10}$(sSFR) (yr$^{-1}$) & & $-9.88\pm0.05$ & $-9.43\pm0.07$ & $-$ \\
12+log(O/H) & & $8.571\pm0.006$ & $8.565\pm0.009$ & $-$ \\
\enddata
\tablecomments{The velocity, velocity dispersion and redshift measurements were calculated from the gas emission lines in the galaxies; for the GC these measurements were performed with the stellar absorption lines.}
\end{deluxetable*}

\subsection{Surface brightness profiles}\label{sec:SB_fits}
The next step of the analysis was to calculate the galaxy masses through fits to the surface brightness profiles of each galaxy.~We first determined the structural parameters of the galaxies, such as size and luminosity, by applying light profile fits to the NGFS $g^\prime$ and $i^\prime$-band images with \textsc{galfit} \citep[v3.0.5,][]{Peng_2010}.~The NGFS data were selected for this measurement over the HST data available since the effects of the seeing smooth out any substructure and asymmetries in the images and allows a better global fit to the data in order to measure the total flux from the system. Additionally, we used \textsc{galfit} to measure the magnitudes as opposed to using the measurements from the aperture photometry since \textsc{galfit} could model the globular cluster as well as the galaxies, and thus better disentangle the light from the three sources.~The first step was to create a postage-stamp image of the target galaxies from the NGFS $g^\prime$ and $i^\prime$-band images and then construct the corresponding segmentation maps using \textsc{SExtractor} \citep{Bertin_1996}.~A bad pixel mask was created from this segmentation map, and was used with \textsc{galfit} to minimize contamination from other sources in the field-of-view.~A point-spread function (PSF) model was also created for each image using PSFEx \citep{Bertin_2011}, which was used in the fits to account for the effect of the seeing.

Due to the proximity of NGC~1396, the light from the outskirts of this foreground galaxy is superimposed upon the target galaxies.~Therefore, to obtain a reliable fit to the target galaxies, this contamination was  quantified and removed to leave a smooth sky background behind the target galaxies. This step was carried out by first modeling NGC\,1396 with \textsc{galfit}, starting with a single S\'ersic model and adding additional profile components until the residual image, created by subtracting the model from the input image, showed only insignificant artifacts.~Having removed the light from the foreground galaxy, the target galaxies were then fitted by \textsc{galfit} directly on the residual image.~The fits to the target galaxies were carried out using S\'ersic profiles, and a single S\'ersic profile was found to sufficiently model each galaxy with residuals near or below the sky noise.The fit to Galaxy~A resulted in a S\'ersic index of 1.7 and 1.4 in the $g^\prime$ and $i^\prime$-bands respectively, while the fit to Galaxy~B required a S\'ersic index of 0.5 in both bands. Galaxy~B was modelled similarly well with a PSF profile, with consistent magnitudes, but in this study we will use the S\'ersic profile for consistency between the two targets.~The foreground globular cluster was also modelled  as a PSF profile at the same time to reduce contamination to the galaxy light from this source.~The physical properties are listed in Table~\ref{tab:kinematics}, including the total apparent magnitudes, the total absolute magnitudes which have been k-corrected, the effective radii in both arcseconds and kpc, and the mean effective surface brightnesses after correction for cosmological dimming.~It should be noted that the errors on the physical parameters are calculated by \textsc{galfit} assuming that the galaxy is well modeled with the number of fit components and that the images contain only Poissonian noise.~Since such a scenario is very rare in reality, the uncertainties presented here for the magnitudes and effective radii should be considered lower limits \citep{Haeussler_2007}.~Compared to non-star-forming dwarf galaxies of similar stellar mass in the Fornax cluster \citep[see][]{Eigenthaler_2018, Ordenes_2018a} our two target dwarfs have $\sim\!3$ mag higher average effective surface brightness levels, owing likely to the ongoing star formation.~Consistently, their optical rest-frame colors put them on the blue sequence at their stellar mass \citep[see~Fig.~6 in][]{Eigenthaler_2018}.

\subsection{Determination of the stellar mass and size}\label{sec:physical_properties}

In order to measure the stellar mass of a galaxy one must obtain an estimate of the stellar populations present. In general, using the stellar absorption lines is a more powerful and reliable method to obtain the stellar population properties than the integrated colours. The spectra for the two galaxies were modeled again using pPXF with the regularization parameter to measure the mass-weighted star-formation histories of the two galaxies. However, as can be seen in the zoom-in of the Mg triplet and Fe lines in Fig.~\ref{fig:fits}, the S/N of the stellar absorption spectra is very low, and so
the results obtained through this method were considered unreliable. Therefore, the mass estimates were instead obtained using the colour information from the NGFS data.

The luminosity-weighted stellar masses were calculated by fitting stellar population synthesis models to the integrated $g^\prime$ and $i^\prime$-band magnitudes of the galaxies obtained from the fits.~The SSP models of \citet{Bruzual_2003} were used with the IMF of \citet{Kroupa_2001}, covering a metallicity range of $0.0001\le Z/Z_\sun\le0.5$. 
The $V$-band mass-to-light ratio was used in conjunction with the  $g^\prime\!-\!i^\prime$ colours since it is the optimal luminescent passband for estimating the wavelength-dependent mass-light ratio from single colours, and reduced the  systematic dependence on the dust extinction, metallicity and star-formation history \citep{Zhang_2017}.~The stellar masses of galaxies~${\cal A}$ and ${\cal B}$ were found to be 10$^{8.8\pm0.3}~\text{M}_\sun$ and 10$^{8.2\pm0.2}~\text{M}_\sun$ respectively, putting them into the dwarf mass regime.~The uncertainties were estimated through a series of Monte Carlo simulations, creating a normal distribution for the magnitudes with the same mean and standard deviation as those measured from the galaxy pair, and taking 1000 random values from within this distribution and repeating the stellar mass calculation outlined above.~The analysis presented in this section was also repeated using the $g^\prime$ and $i^\prime$-band images created from the MUSE datacube, with consistent results despite the residual structure in the background of those images due to the slicers and channels within the MUSE light path.

\begin{figure}
\begin{center}
\includegraphics[scale=0.99,angle=0,width=0.99\linewidth]{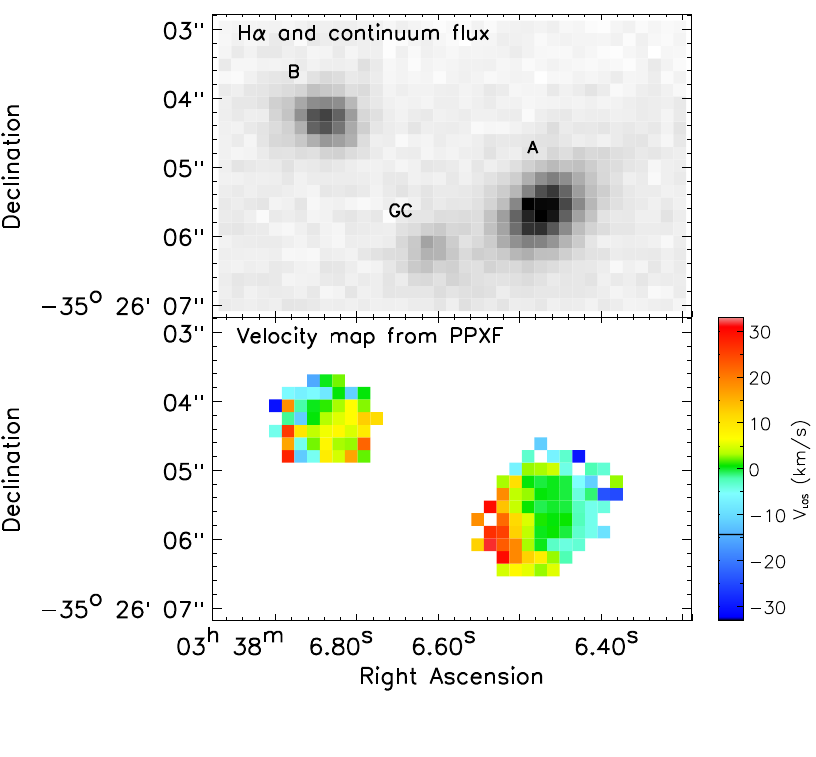}
\caption{A white-light image of the dwarf pair from the MUSE datacube covering the H$\alpha$ emission line and the continuum (\textit{top}), with the gas velocity map created by fitting all the emission lines in the spectrum with pPXF (\textit{bottom}). 
\label{fig:rotation}}
\end{center}
\end{figure}

\subsection{Determination of the gas rotation}\label{sec:rotation}
Disturbed kinematics within a galaxy can be a signature of a recent interaction with a neighbour.~The strong and sharp emission lines within these galaxies were, therefore, used to study the gas rotation and look for evidence of an interaction.~A postage-stamp datacube was extracted from the MUSE observations centered on the dwarf-dwarf pair, and the emission line velocity was measured for each spaxel using pPXF with only the emission line templates to fit all the emission features present in each spectrum. A low-order polynomial was included in the fit to represent the continuum. The resultant velocity map is presented in Figure~\ref{fig:rotation}, alongside a white-light image of the system showing the H$\alpha$ emission and continuum.~The velocity map has been cleaned to remove the results from poor fits and from spaxels with too low flux to determine a reliable velocity.~Additionally, the spaxels whose light is dominated by the foreground globular cluster have also been omitted. This map reveals that  Galaxy~${\cal A}$ is rotating, with a global kinematic position angle of $111\pm10$\textdegree\ as measured by the method described in Appendix~C of \citet{Krajnovic_2006}, which is roughly aligned with the major axis. Rotation in Galaxy~${\cal B}$ is much more tentative due to its smaller size on the sky, with the same analysis resulting in a kinematic position angle of $-140\pm44$\textdegree.

In order to confirm  these results, the emission velocity in each spaxel was measured again using cross correlation for the brightest emission lines in the spectrum (H$\alpha$, [O{\sc iii}]$_{5007}$ and [O{\sc ii}]$_{3727/3729}$). Cross correlation was applied to every spaxel in both galaxies, using the same emission features in the brightest spaxel of Galaxy~${\cal A}$ as the template since measuring offsets relative to this spaxel would emphasize any significant offsets in the gas emission between the galaxies. The relative pixel shifts measured by this technique showed the same rotation in Galaxy~${\cal A}$, confirming that it is rotating, while no clear rotation was detected in Galaxy~${\cal B}$.~The rotation in Galaxy~${\cal A}$, therefore, reflects that the gaseous orbits within this galaxy have not been significantly disrupted by the proximity of Galaxy~${\cal B}$. According to simulations by \citet{Hung_2016}  of galaxy mergers with mass ratios of 1:1 and 1:4, kinematic merger signatures become most apparent after the second passage, and that $\sim20-40\%$ of galaxies are not identified as mergers before the strong interaction phase. Thus, the ordered kinematics seen in Galaxy~${\cal A}$ suggests that these galaxies are still in the early stages of interacting, before the second passage and strong interaction phase. The tidal tail in Galaxy~${\cal A}$ that can be seen in Fig.~\ref{fig:images} may therefore be indicative of an earlier merger or of disruption caused during the first passage.

\subsection{Determination of the ionization mechanism}\label{sec:BPT}
The presence of emission lines in both galaxies indicates that the gas is ionized. While the ionization mechanism is most likely ongoing star-formation, the recent discovery of 136 dwarf galaxies at $z<0.055$ with optical emission line signatures of AGN by \citet{Reines_2013} mean that we cannot rule out this scenario. Furthermore, the presence of emission lines could also be explained by feedback-driven shocks \citep{Calzetti_2004,Thuan_2005, Hong_2013, Davies_2017}.~In order to determine the ionization mechanism, the ratios of the line fluxes can be plotted onto the BPT~diagram \citep{Baldwin_1981} to distinguish between star-forming/H{\sc ii} regions, and AGN (Seyfert and LINERs)  and shocks \citep[e.g.][]{Schawinski_2007}.

\begin{figure*}
\begin{center}
\includegraphics[angle=0,width=0.7\linewidth]{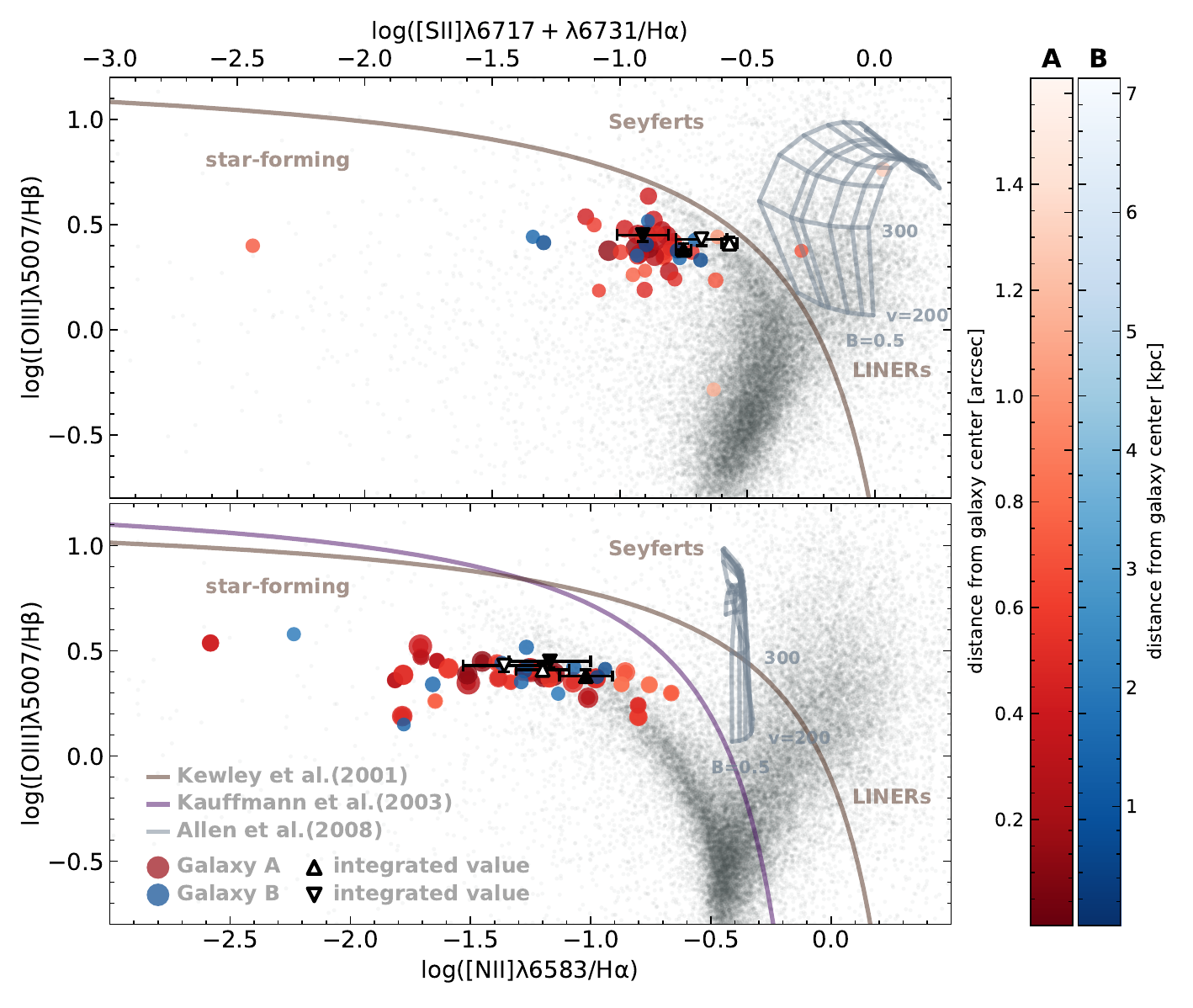}
\caption{The BPT diagrams for each spaxel covered by Galaxy~${\cal A}$ (red) and ${\cal B}$ (blue), where the size of the point represents the S/N of that measurement and the shading reflects the distance from the centre of the galaxy. The grey points represent the measurements from the MPA-JHU SDSS DR7
 galaxies for comparison, and the upward and downward pointing triangles with error bars represent the measurements from the integrated spectra for Galaxy~${\cal A}$ and ${\cal B}$ (filled) and the best fit to the gas emission lines from that spectrum (hollow), respectively. The shock-ionization model grids of \citet{Allen_2008} are overlaid for reference, showing the predicted line ratios for shocks with velocities of 200-1000~km/s (horizontal lines), magnetic fields of strengths 0.5, 1, 2, 4, 5 and 10~$\micro$G (vertical lines) and for solar abundances from \citet{Dopita_2005}.
\label{fig:BPT}}
\end{center}
\end{figure*}

Emission maps were created for the H$\beta$, [O{\sc iii}]$_{\lambda5007}$, H$\alpha$,  [N{\sc ii}]$_{\lambda6583}$, [S{\sc ii}]$_{\lambda6717}$ and [S{\sc ii}]$_{\lambda6731}$ lines by taking the flux over that line for each spaxel, defining a wide enough wavelength range such that the wings of the line do not move outside of the measured range due to the rotational velocity, and narrow enough to minimize blending with neighbouring emission features.~The continuum level was measured from regions to one side or on both sides of each emission line that was devoid of strong features, and was subtracted from each spaxel in the map.~The flux ratios [O{\sc iii}]$_{\lambda5007}$/H$\beta$, [N{\sc ii}]$_{\lambda6583}$/H$\alpha$ and  [S{\sc ii}]$_{\lambda6717+6731}$/H$\alpha$ were then calculated for each spaxel to determine it's location on the BPT~diagram.~The results are shown in Figure~\ref{fig:BPT} for each spaxel over the two galaxies, with Galaxies~${\cal A}$ and ${\cal B}$ marked in red and blue, respectively, and with the size and shading of each point representing the S/N of that spaxel and its distance from the centre of the galaxy as determined by the brightest spaxel in the optical bands.~Measurements of other emission-line galaxies from the MPA-JHU SDSS-DR7 catalog\footnote{https://wwwmpa.mpa-garching.mpg.de/SDSS/DR7/}, the curves of \citet{Kewley_2001} and \citet{Kauffmann_2003b}, and the MAPPINGS shock-ionization models of \citet{Allen_2008} are included for reference. The results clearly show that both galaxies lie in the star-forming regime of the BPT diagram and show a larger spread in the horizontal direction of both plots compared to the vertical direction.~While this trend may represent a gradient or fluctuations in the line strengths over the galaxies, it is more likely to reflect the noise in the measurements due to the relative weakness of the [N{\sc ii}]$_{\lambda6583}$ and  [S{\sc ii}]$_{\lambda6717+6731}$ lines compared to the H$\alpha$ feature. 

The MAPPINGS shock-ionization models plotted in Fig.~\ref{fig:BPT} have been calculated for a solar abundance based on \citet{Dopita_2005}, which was found to be closest model to the abundance measured in the dwarf-dwarf pair (see Section~\ref{sec:gas_met}). However, the metallicity of the dwarf-dwarf pair varies depending on the calibration used, and the shock-ionization models will shift to the left of the BPT diagrams for lower metallicities. The MAPPINGS models have been created for a range of abundances, of with the sub-solar LMC abundance also falls within the range of measurements described in Section~\ref{sec:gas_met}. However, plotting these models onto Fig.~\ref{fig:BPT} showed that the results are unaffected for the range of metallicities measured for this system, and the LMC models have been omitted from Fig.~\ref{fig:BPT} for clarity.

It should be noted that the H$\beta$ and H$\alpha$ emission lines lie at the same wavelength as the corresponding absorption features.~Since the S/N of the spectra from individual spaxels was too low to allow a reliable fit to the stellar absorption lines with pPXF, these flux measurements are therefore not absorption corrected.~This effect leads to a small systematic offset in the BPT measurements such that these emission line fluxes are underestimated, and thus overestimating the [O{\sc iii}]$_{\lambda5007}$/H$\beta$, [N{\sc ii}]$_{\lambda6583}$/H$\alpha$ and [S{\sc ii}]$_{\lambda6717+6731}$/H$\alpha$ flux ratios. In order to quantify this effect, the line strengths were measured again in the integrated spectra for each galaxy and from the best fit the  emission lines obtained from pPXF in Section~\ref{sec:redshift}. The results are plotted in Figure~\ref{fig:BPT} as filled and hollow black triangles respectively. It can be seen that the offset in the [O{\sc iii}]$_{\lambda5007}$/H$\beta$ ratio is very small for both galaxies, indicating that the emission lines are significantly stronger than the corresponding absorption features, and thus the offset in the emission line strength is within the uncertainties. 

Similarly small offsets are seen in the [N{\sc ii}]$_{\lambda6583}$/H$\alpha$ line ratios, while an offset is seen in the [S{\sc ii}]$_{\lambda6717+6731}$/H$\alpha$ line ratio measured from the emission-only spectrum compared to both the individual pixels and the integrated spectra for both galaxies. While these measurements are still within the scatter from the measurements of the individual spaxels, the offset in the [S{\sc ii}]$_{\lambda6717+6731}$/H$\alpha$ line ratio is likely an effect of the low S/N of the spectra. However, taking the scatter in the results from the individual spaxels as a guide to the range within which the true values lie, one can see the emission line characteristics  put both galaxies firmly in the star-forming regime.

\subsection{Determination of the star-formation rate}\label{sec:SFR}
As evident from Figure~\ref{fig:BPT}, both galaxies are actively star forming, and the MUSE spectra cover the redshifted wavelength of H$\alpha$, a key star-formation rate (SFR) indicator \citep[e.g.][]{Kennicutt_1998, Kewley_2002}.~We can therefore use this emission to compute the present SFR.~Prior to this calculation, the H$\alpha$ flux must be measured after subtracting the absorption contribution, and then corrected for dust reddening.~The emission spectrum obtained for each galaxy was taken from the pPXF fits to the integrated spectra shown in Figure~\ref{fig:fits}, and the H$\alpha$ and H$\beta$ fluxes ($F_{\text{H}\alpha}$, $F_{\text{H}\beta}$) measured by fitting a Gaussian profile to these lines.~The uncertainty in the H$\alpha$ and H$\beta$ fluxes was taken as the standard deviation in the residuals over the same wavelength region, which should also include the effects of the poor fits to the absorption features by pPXF.

\begin{figure*}
\begin{center}
\includegraphics[angle=0,width=0.8\linewidth]{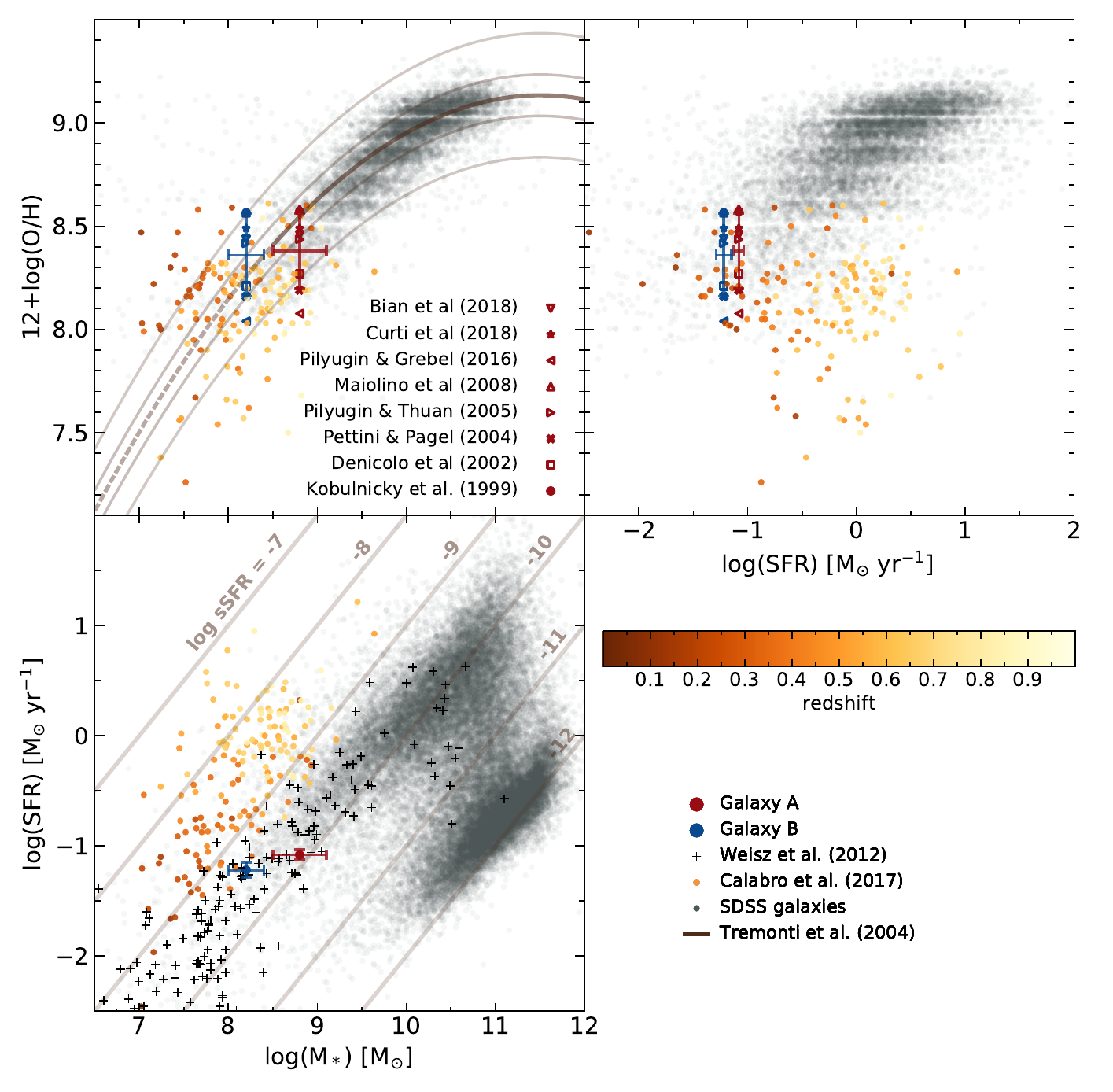}
\caption{The relationship between the stellar mass, metallicity and SFR for the dwarf galaxy pair, superimposed upon the results for star-forming dwarf galaxies from \citet{Calabro_2017} (dots, colour-coded according to their redshift), galaxies from the MPA-JHU SDSS DR7 \citep[black dots,][]{Brinchmann_2004, Tremonti_2004}, and the mass-SFR relation of \citet{Weisz_2012} (crosses) for local star-forming dwarf galaxies. The mass-metallicity relation of \citet{Tremonti_2004} for star-forming SDSS galaxies in the mass range $8.5<\log_{10}({\cal M}_*)<10.5$ and $z\!\lesssim\!0.1$ has been over plotted for reference (dark shaded curve), with the dashed curve at $\log_{10}({\cal M}_*)<8.5$ showing the extrapolation to lower masses, and the light-shaded curves representing the 1 and 3\,$\sigma$ offsets. The lines of constant sSFR have also been plotted for reference on the mass-SFR plot. The hollow symbols represent the different metallicity measurements when using calibration corefficients from the literature for various line ratios, and the error bars indicate the mean and standard deviation of these measurements (see Section~\ref{sec:gas_met}). 
\label{fig:met}}
\end{center}
\end{figure*}

The H$\alpha$ dust attenuation correction was applied using the foreground dust screen approximation.~We assumed an intrinsic Balmer ratio of 2.86 for case-B recombination \citep{Osterbrock_1989} at an electron temperature $T_e\!=\!10\,000$\,K and density $n_e\!=\!100$\,cm$^{-3}$ \citep{Hummer_1987}.~The dust attenuation on the H$\alpha$ line was then corrected using the extinction map of \citet{Schlegel_1998} combined with the Galactic extinction law of \citet{Fitzpatrick_2007}:
\begin{equation}   
A({H}\alpha) = 1.754 \times C({{H}\beta}),
\label{eq:dust_attenuation}
\end{equation}
where the Balmer decrement was defined as
\begin{equation}
C({\text{H}\beta}) = \frac{log(2.86)-log\Big[\frac{F(H\alpha)}{F(H\beta)}\Big]_{obs}}{f(H\alpha)},
\label{eq:C_lambda}
\end{equation}
and $f(H\alpha)=-0.297$ is the reddening function relative to $H\beta$. When applied to the \citet{Calzetti_2000} starburst reddening curve,
\begin{equation}
k(\lambda) =A_{\lambda}/E(B-V),   
\label{eq:EBV}
\end{equation} 
where $k(\lambda)$ is the effective attenuation and has a value of $k(\lambda)=k(\text{H}\alpha)=3.33\pm0.80$ at the wavelength of $H\alpha$, this dust attenuation corresponds to a colour excess of $E(B-V)=0.06\pm0.02$~mag and $0.10\pm0.03$~mag for Galaxies~${\cal A}$ and ${\cal B}$ respectively.
While the dust attenuation correction should be applied to the H$\alpha$ flux in every spaxel before calculating the integrated flux for each galaxy, the low S/N of the spectra in individual spaxels is likely to introduce noisy corrections.~\citet{Catalan_2015} compared the dust attenuation measurements for spectra in individual spaxels of a CALIFA datacube with that measured from the integrated spectrum, and found that the difference was less than 1\%. They conclude that it is safe to use the ratio of the H$\alpha$/H$\beta$ fluxes derived from an integrated spectrum to correct the H$\alpha$ flux of that galaxy for each spaxel.~We follow this approach.

The SFR is then calculated in units of M$_{\sun}$ yr$^{-1}$ using the absorption and continuum-corrected H$\alpha$ flux in the relation
\begin{equation}
{\rm SFR}({H}\alpha)=7.9 \times 10^{-42} L({H}\alpha)
\label{eq:mass_dyn}
\end{equation}
from \citep{Kennicutt_1998}.~The measurements for the SFR for both galaxies are presented in Table~\ref{tab:kinematics}, and the SFRs are plotted against the stellar mass of the galaxy in Figure~\ref{fig:met}.~We compare our sample with the results from \citet{Calabro_2017}, who measure the properties of star-forming dwarf galaxies with stellar masses of $\log_{10}({\cal M}_*/M_\sun)\!<\!10$ over the redshift range $0.1\!<\!z\!<\!0.9$, and with \citet{Weisz_2012}, who study the properties of a sample of local, star-forming dwarf galaxies, and with the MPA-JHU SDSS DR7 galaxy sample, with SFRs from \citet{Brinchmann_2004} and masses derived from photometry following the philosophy of \citet{Kauffmann_2003}.~It can be seen that the SFRs for both dwarf galaxies are consistent with similar mass star-forming SDSS galaxies and dwarf galaxies in the \citet{Weisz_2012} sample. The \citet{Calabro_2017} sample however shows an offset towards higher SFRs at a given stellar mass, although the SFRs are consistent when only the redshift is considered. This offset is attributed to a selection bias in that sample imposed by selecting galaxies using a S/N cut on the hydrogen recombination line strength, which limits the sample to galaxies with brighter H$\alpha$ features, and thus higher specific star-formation rates (sSFR), than a similar-mass continuum selected sample. A comparison of the sSFRs of Galaxies~${\cal A}$ and ${\cal B}$ of $\log_{10}(sSFR)=-9.88\pm0.05$~yr$^{-1}$ and $-9.43\pm0.07$~yr$^{-1}$ respectively shows that they are lower than the median values of $-8.3$~yr$^{-1}$ and $-8.6$~yr$^{-1}$ for extreme and non-extreme emission line galaxies in the \citet{Calabro_2017} sample. \citet{Calabro_2017} go on to explain that these galaxies also have low metallicities, unusually high gas mass fractions, and that they either have very inefficient star formation or are very young and still assembling their stars. 

With a SFR of less than $\sim\!0.1\,M_{\sun}$\,yr$^{-1}$ , this system shows a lower SFR than the isolated dwarf pairs in the \textsc{TiNy Titans} survey \citep{Stierwalt_2015}.~Instead, it resembles the typical SFR seen for the same projected separation in their isolated and non-isolated single dwarfs that have been matched for ${\cal M}_*$, redshift, local density and isolation.~Consequently, it is likely that these galaxies form a non-interacting pair, or are in the earliest stages of merging, such that the SFR has not yet experienced a significant enhancement, which coincides with the second passage in the simulations of \citet{Hung_2016}. These scenarios are consistent with the seemingly undisturbed rotation pattern in Galaxy~${\cal A}$, and the second scenario may also explain the disturbed morphology in Galaxy~${\cal A}$ if the galaxies are between the first and second passages.

\subsection{Determination of the gas metallicity}\label{sec:gas_met}
The gas metallicities of the two dwarf galaxies can provide some indications towards the origin of the two galaxies and the current state of their evolution.~For example, if they have formed from the same gas cloud, their gas metallicities may be expected to be similar, whereas if they have formed independently from separate sources of gas, their gas properties should also be different.~Therefore, to determine which scenario is the more likely, we derive and compare the metallicities of the gas in both galaxies.

The gas-phase metallicity is measured using the oxygen abundance as a tracer alongside the $R_{23}$ intensity ratio,
\begin{equation}
R_{23}=\log \bigg(\frac{{\rm [O\text{\sc ii}]}_{\lambda3727} + {\rm [O\text{\sc iii}]}_{\lambda4959}  + {\rm [O\text{\sc iii}]}_{\lambda5007} }{\text{H}\beta}  \bigg) \equiv X,
\label{eq:R23}
\end{equation}
\citep{Pagel_1979}. Since $R_{23}$ is sensitive to the ionization state of the gas, the $O_{32}$ ratio,
\begin{equation}
O_{32}=\log \bigg(\frac{{\rm [O\text{\sc iii}]}_{\lambda4959}  + {\rm [O\text{\sc iii}]}_{\lambda5007} }{{\rm [O\text{\sc ii}]}_{\lambda3727} }  \bigg) \equiv Y,
\label{eq:O23}
\end{equation}
was calculated as a correction.~Particular care has to be taken when using the $R_{23}$ ratio, as most values can be degenerate and correspond to both a low-metallicity estimate (lower branch) and a high-metallicity estimate (upper branch).~This is equivalent to a horizontal cut in the upper section of both panels in Figure~\ref{fig:BPT} .~In order to distinguish between these two branches, one would normally use the ${\rm [N\text{\sc ii}]}/{\rm [O\text{\sc ii}]}$ ratio \citep{Kewley_2008}.~However, in these galaxies, the [N{\sc ii}] line is very weak, and both galaxies were found to have ratios of $\log({\rm [N\text{\sc ii}]}/{\rm [O\text{\sc ii}]})\!\simeq\!-1.2$, making the differentiation between the upper and lower branches difficult. Instead, we use the ratio of ${\rm [O\text{\sc iii}]}_{\lambda5007}/{\rm [O\text{\sc ii}]}_{\lambda3727}\!=\!2.0$, which corresponds to $12+\log({\rm O/H})\simeq8.0$ \citep{Nagao_2006}, to break the degeneracy. We measure this line strength ratio to be $\sim\!0.9$ for both galaxies, putting them on the upper branch. Thus the oxygen abundance is calculated using
\begin{equation}
\begin{split}
12+\log(O/H)=9.061 - 0.2X - 0.237X^2 - 0.305X^3 \\
- 0.0238X^3 - Y(0.0047 - 0.00221X - 0.102X^2 \\
- 0.0817X^3 - 0.00717X^4)
\end{split}
\label{eq:O/H}
\end{equation}
\citep{Kobulnicky_1999}.~Both galaxies are found to have very similar metallicities, with $12+\log({\rm O/H})\!=\!8.571\pm0.006$ and $8.564\pm0.009$ for Galaxies~${\cal A}$ and ${\cal B}$, respectively.~The uncertainties are derived from the standard deviation in the residual spectrum over the same wavelength range over which the line fluxes were measured.~Figure~\ref{fig:met} (top left panel) demonstrates that the dwarf galaxies fall close to the average mass-metallicity relation of \citet{Tremonti_2004}, with Galaxies~${\cal A}$ and ${\cal B}$ located within 1 and 3\,$\sigma$ of their trend line.~Furthermore, in the SFR-metallicity plot in Figure~\ref{fig:met} (top right panel), both galaxies show consistent results with the SDSS sample. The galaxies however are offset to higher metallicities than the star-forming dwarf galaxy sample of \citet{Calabro_2017}, which is due to the selection bias on that sample towards lower-metallicity galaxies (see Section.~\ref{sec:SFR}). 

Many different calibrations exist for measuring the metallicity of a star-forming galaxy using different line ratios, and it is unlikely that the small errors associated to the metallicity measurements listed above represent the true dispersion. Consequently, a series of additional measurements were made using calibrators from the literature for different emission lines. The calibrations included those of \citet{Pilyugin_2005}, \citet{Maiolino_2008}, \citet{Curti_2017} and \citet{Bian_2018}, which use the  $R_{23}$ line ratio;  \citet{Pilyugin_2016}, which uses the  [NII]/H$\beta$ line ratio in addition to $R_{23}$;  \citet{Denicolo_2002}, which uses the [NII]/H$\alpha$ line ratio; and finally \citet{Pettini_2004}, which uses the [OIII]$_{5007}$/H$\beta$ and [NII]/H$\alpha$ line ratios. The results for these measurements are given in Fig.~\ref{fig:met}, and the error bars overlaid upon these results represent the mean and standard deviation. With this distribution, both galaxies still show metallicities that lie within 3\,$\sigma$ of the mass-metallicity relation of \citet{Tremonti_2004}.

\section{Discussion and Conclusions}\label{sec:conclusions}
Through the use of deep VLT/MUSE IFU spectroscopic and NGFS and HST imaging data, we detected a pair of galaxies that appear close on sky and with emission features at similar wavelengths.~The measurements of their gas kinematics from the emission lines reveal that both galaxies lie at the same redshift, $z\!=\!0.30$ with radial velocities of 76967 and 76975~km/s, while estimates of their stellar masses from their $g^\prime$ and $i^\prime$-band magnitudes show that they are both dwarf galaxies with $\log({\cal M}_*/M_{\odot})\!\lesssim\!9$.~Previous studies by \citet{Hilker_1999} and \citet{Drinkwater_2000} have found evidence of large-scale filamentary structure extending out to $z\!\simeq\!0.3$ behind the Fornax cluster. With the deep MUSE data, we detected background targets up to $z\sim1.3$ with no galaxies within the field-of-view of the IFU datacube with strong emission lines at the same redshift as the dwarf-dwarf pair. Using spectroscopic redshifts from the literature for all galaxies within a volume of radius 1.5~Mpc on sky and brighter than an apparent magnitude of 10 (FCSS) and 13~mag (2DF) in the \textit{b$_j$} band, the nearest massive neighbour was found to lie at a distance of 38~Mpc, and so is not considered to be a companion. Determination of photometric redshifts of all fainter objects within this volume was attempted using the NGFS $u^\prime$, $g^\prime$, $i^\prime$ and $K_s$-band images, but a comparison with the spectroscopic redshifts from the MUSE field proved that the results were unreliable due to the small number and poor distribution of filters in wavelength space. Therefore, with the data available, we have been unable to determine whether these galaxies lie isolated in the field or are members of a group of low-mass galaxies.

The fluxes of the prominent hydrogen, nitrogen, oxygen, and sulphur emission lines were measured, and the flux ratios plotted onto the BPT diagram to determine the ionization mechanism.~This plot reveals that the gas ionization in both galaxies is driven by ongoing star formation, rather than by AGN activity or shocks.~The SFR calculated from the H$\alpha$-emission flux shows that both galaxies lie on the SFR main sequence for their corresponding stellar masses.~Furthermore, their SFRs are consistent with  local isolated dwarf galaxies of similar masses, and with dwarf-dwarf pairs with larger separations ($>\!200$kpc) in the nearby universe \citep{Stierwalt_2015}.~Studies of local ongoing dwarf-dwarf mergers have found that the SFR in such interactions are typically enhanced by a factor of $\sim2.3$ compared to unpaired dwarf galaxies of similar masses  \citep{Stierwalt_2015,Privon_2017}, which does not appear to be the case for the system presented in this paper.~Additionally, the analysis of the gas velocities across both galaxies reveals that Galaxy~${\cal A}$ shows no evidence of disturbed gas orbits, instead showing ordered rotation with an amplitude of $\pm30$km/s. Simulations by \citet{Hung_2016} showed that the SFR is enhanced significantly and that kinematical signatures of mergers become most apparent after the second approach. Therefore, we believe that this system is still in the early stages of interaction, likely with both galaxies on their first or second approach such that they have not yet disrupted one another and triggered a period of enhanced star formation. 

In the HST image of the system in Figure~\ref{fig:images}, Galaxy~${\cal A}$ appears morphologically disturbed, with a tail extending towards the north-west and bright points which may reflect localized enhanced star-forming regions.~These signs are typical signatures of a merger or interaction, but with the lack of evidence that Galaxy~${\cal A}$ and ${\cal B}$ are currently interacting, we conclude that  Galaxy~${\cal A}$ has either separately undergone a merger with another galaxy that was recent enough that its morphology is still disturbed, but long enough ago that the SFR is no longer enhanced, or it contains a  regions of bright star formation that mimic  the asymmetric appearance of an otherwise normal gas disc. \citet{Bekki_2008} determined that the period of enhanced SFR during a dwarf-galaxy merger is $\sim\!0.1$\,Gyr, and so with the data available we can only determine that any merger involving Galaxy~${\cal A}$ likely happened more than 100~million years ago. The gas-kinematics map for galaxy~${\cal A}$ shows no evidence of these tidal tails, suggesting that the tails themselves contain too little gas to detect with the MUSE spectra, and that the majority of the gas has either maintained its rotation or has settled back into a disc after the merger. 

Due to the faintness of the targets and the low S/N of their spectra, it was not possible to carry out a reliable stellar populations analysis using the absorption features to determine the ages and star-formation histories of the galaxies. Instead, the gas metallicities were calculated from the emission lines to give an indication of the origin of the two galaxies. The gas metallicities of both galaxies are consistent, indicating that they likely formed from the same source material as opposed to having formed independently and happen to be near each other.~While the similarity in the gas metallicities may reflect that one galaxy has accreted gas from the other, the un-enhanced SFR and undisturbed gas rotation in Galaxy~${\cal A}$ contradict such a scenario.

Through the analysis presented in this paper, we conclude that these dwarf galaxies are part of a non-interacting pair or are in the early stages of an interaction, on the first or second approach. We find no evidence of more massive neighbours, and conclude that they  lie either in the field or within a group of low-mass ($\log{\cal M}_*\!\lesssim\!8$) galaxies.~Such groups of dwarf galaxies have so far only been detected in the local Universe and out to $z\!\simeq\!0.1$ in small numbers \citep[e.g.][]{Tully_2006, Koch_2015, Stierwalt_2017}, and so the detection of a similar group at $z\!=\!0.3$ provides a new opportunity to explore the properties of dwarf galaxy groups and galaxy-galaxy interactions in the dwarf mass regime as a function of redshift.

\acknowledgments

We would like to thank the anonymous referee for their suggestions, which helped to improve this paper. This study was based on data obtained from the ESO Science Archive Facility under request number P094.B-0895 (PI: Lisker).~Based on observations made with the NASA/ESA Hubble Space Telescope, and obtained from the Hubble Legacy Archive proposal ID:10129 (PI: Puzia), which is a collaboration between the Space Telescope Science Institute (STScI/NASA), the Space Telescope European Coordinating Facility (ST-ECF/ESA) and the Canadian Astronomy Data Centre (CADC/NRC/CSA).~This project used data obtained with the Dark Energy Camera (DECam), which  was constructed by the Dark Energy Survey (DES) collaboration.~

This project is supported by FONDECYT Regular Project No.\,1161817 and the BASAL Center for Astrophysics and Associated Technologies (PFB-06). E.J.J. acknowledges support from FONDECYT Postdoctoral Fellowship Project No.\,3180557, Y.O.-B.\ acknowledges financial support through CONICYT-Chile (grant CONICYT-PCHA/Doctorado Nacional No.~2014-21140651) and the DAAD through the PUC-HD Graduate Exchange Fellowship.~T.H.P.~and A.L.~gratefully acknowledge ECOS-Sud/CONICYT project C15U02.~P.E.~acknowledges support from the Chinese Academy of Sciences (CAS) through CAS-CONICYT Postdoctoral Fellowship CAS150023 administered by the CAS South America Center for Astronomy (CASSACA) in Santiago, Chile. M.A.T. is supported by the Gemini Observatory, which is operated by the Association of Universities for Research in Astronomy, Inc., on behalf of the international Gemini partnership of Argentina, Brazil, Canada, Chile, the Republic of Korea, and the United States of America.YR acknowledges supports from CAS-CONICYT postdoctoral fellowship No.\,16004 and NSFC grant No.\,11703037.

\end{document}